\documentclass[prb,twocolumn]{revtex4}
\usepackage{graphicx}
\usepackage{subfigure}
\usepackage{color}
\usepackage{amsmath}
\usepackage{amssymb}
\usepackage{wasysym}
\usepackage{bm}
\usepackage{placeins}
\usepackage[bookmarks=false]{hyperref}
\usepackage{slashed}
\newcommand{\be}{\begin{equation}}
\newcommand{\ee}{\end{equation}}
\newcommand{\ba}{\begin{eqnarray}}
\newcommand{\ea}{\end{eqnarray}}

\makeatletter
\newcommand{\crossout}[1]{%
  \begingroup
  \sbox\z@{#1}%
  \dimen\z@=\wd\z@
  \dimen\tw@=\ht\z@
  \dimen\z@=.99626\dimen\z@   % get big points
  \dimen\tw@=.99626\dimen\tw@ % get big points
  \edef\co@wd{\strip@pt\dimen\z@}%  just the number
  \edef\co@ht{\strip@pt\dimen\tw@}% just the number
  \leavevmode
  \rlap{\pdfliteral{q 1 J 0.4 w 0 0 m \co@wd\space \co@ht\space l S Q}}%
  \rlap{\pdfliteral{q 1 J 0.4 w 0 \co@ht\space m \co@wd\space 0 l S Q}}%
  #1%
  \endgroup
}
\makeatother

\begin{document}

\title{Phases of a phenomenological model of twisted bilayer graphene}
\author{  {$\rm J. \ F. \ Dodaro^1$}, {$\rm \ S. \ A. \ Kivelson^{1}$}, {$\rm \ Y. \ Schattner^{1,2}$}, {$\rm \ X. \ Q. \ Sun^{1}$}, {$\rm and \  C. \ Wang^{1}$}  }
\address{1) Department of Physics, Stanford University, Stanford, CA 94305}
\address{2) Stanford Institute for Materials and Energy Sciences, SLAC National Accelerator Laboratory and Stanford University, Menlo Park, CA 94025}

\begin{abstract}
    
{We propose a lattice scale two-band generalized Hubbard model as a caricature of the electronic structure of twisted bilayer graphene.  Various possible broken symmetry phases can arise, including a nematic phase (which is a form of orbital ferromagnet) and an orbital-triplet spin-singlet superconducting phase.  Concerning the mechanism of superconductivity -- we propose an analogy with superconductivity in alkali-doped C$_{60}$ in which a violation of Hund's first rule plays a central role. }
\end{abstract}

\maketitle

\section{Introduction}

Twisted bilayer graphene makes a moir\'e pattern which defines an approximate triangular lattice with a large unit cell.  Various band-structure calculations show that for certain twist angles, there are remarkably weakly dispersing bands near the Fermi surface.\cite{santos_2007, macdonald_2011}  In particular, the bands immediately above and below the charge neutrality point are thought to be approximately doubly degenerate;  experiments\cite{jarillo-INS_2018, jarillo-SC_2018} show that when one of these bands is filled with four holes  per unit cell, $\overline n_h=4$ ($\overline n_h$ denotes number of holes per unit cell, relative to charge neutrality), there is behavior suggestive of a band insulator, while  for the half-filled band, $\overline n_h=2$, there is what appears to be an emergent insulating state at low temperatures ($T < 4$ K) -- possibly with a broken symmetry.   
(Similar behavior occurs when electrons are added into these bands, $\overline n_h <0$ in our convention.)  As a function of $\overline n_h$, there is a pronounced peak in the resistivity at the charge-neutrality point, $\overline n_h=0$, where the resistivity is a mildly increasing function of decreasing $T$ but does not show any clear tendency to diverge. % reaching a value of around $4h/e^2$ by $T\approx 1K$.  
For hole concentrations, $0 < \overline n_h < 2$ and $2 < \overline n_h < 4$, there appear two superconducting domes with a maximal $T_c \approx 1.7$ K.  (Superconductivity has not yet been detected for $\overline n_h <0$.)

There are many complexities associated with the microscopic physics of this system;  %it has been suggested that
it may be impossible to define localized Wannier functions associated with the two bands just %above
below the charge neutrality point without also including the two %below
above the neutrality point.\cite{senthil_2018,fu_2018} Moreover, unless the twist angle is precisely commensurate, the system in question is in truth a quasi-crystal, not a crystal at all.  
Nonetheless, to simplify the problem, we propose to study a two orbital tight-binding model on the triangular lattice with predominantly on-site interactions, similar to the model introduced in Ref. \onlinecite{xu_2018}, as a caricature of the physical problem.
Note that by construction, our model is a band insulator at the charge neutrality point $\overline n_h=0$, in apparent conflict with experiment, nor can it describe fillings $\overline n_h< 0$.  However, it is our hope that it is sufficient to shed light on the problem in a range of $\overline n_h$ near $\overline n_h\sim 2$.

In thinking about the interactions that enter the model, we should take into account the fact that these are effective interactions.  In particular, since the typical graphene phonon has an energy $\hbar\omega_{\vec q} \sim 200$ meV,\cite{ferrari_2006,ando_2007} while the flat bands in twisted graphene are thought to have band-widths of order 10 meV, \cite{jarillo-INS_2018} the system is in an anti-adiabatic limit in  which it is reasonable to integrate out the phonons. This tends to lower the on-site repulsion between electrons and (from the ``dynamical Jahn-Teller'' effect) produce violations of Hund's rules.
Even with purely repulsive microscopic interactions,  correlation effects associated with integrating out high energy electronic degrees of freedom can also lead to a  reduction of the on-site repulsion and a violation of Hund's rule, as has been shown\cite{jiangandme} for the $t-J$ model on a truncated icosahedron (C$_{60}$). 

With these considerations in mind, a number of broken symmetry phases are possible.  From a weak-coupling perspective, the lack of any Fermi surface nesting on the triangular lattice reduces the susceptibility to translation symmetry breaking.  From a complementary strong-coupling perspective, the triangular lattice is geometrically frustrated.  We will thus focus on states that leave translation symmetry unbroken.\footnote{While these considerations are valid on the triangular lattice, translation symmetry breaking is very natural on a hexagonal and should also be considered in comparing with the physical system.}  Specifically, we have found a regime in which there is an orbital ferromagnetic phase for a range of  $\overline{n}_h$ that extends asymmetrically about $\overline{n}_h=2$.  Depending upon how we identify operators in the model with physical observables, the orbital ferromagnetism can be associated with various patterns of point-group symmetry breaking; for instance it can represent nematic order.  Moreover, the system can be insulating at $\overline{n}_h=2$ if the order is sufficiently strong. Away from $\overline{n}_h=2$ we find two superconducting domes with on-site, spin-singlet, orbital-triplet pairing.

\section{A lattice-scale model}

We define a minimal lattice model with appropriate symmetries and the fewest degrees of freedom needed to account for the salient features summarized above.  We consider a Hubbard-like model on a triangular lattice with a Wannier function that transforms according to a two-dimensional irreducible representation of the point group.   One could think of this as corresponding to  $d_{xy}$ and  $d_{x^2-y^2}$ orbitals split from the other (filled or empty) d-orbitals by crystal field effects, or alternatively as  $p_x$ and $p_y$ orbitals.  At a formal level, these two cases behave similarly, but their different symmetries have different implications for the nature of broken symmetry phases.  To be concrete, we will assume the d-orbital case.

We thus introduce a pseudo-spin index $\tau$, such that $\tau = \pm 1$ corresponds to a $d\pm id$ combination of these orbitals.  We neglect spin-orbit coupling, so there is an $SU(2)$ spin rotational symmetry, and reflecting an assumed conservation of pseudo-spin and time-reversal symmetry  there is an additional orbital $U(1)\times Z_2$ symmetry.  Physically, this $U(1)$ symmetry is related to spatial rotations, and correspondingly a more realistic band-structure would include $\tau$-dependent dispersion relations that would  break the $U(1)$ symmetry to $C_6$;  for the most part we will ignore this for simplicity.  We thus consider the two-orbital model
\be
H=H_0+ H_{int} +H^\prime_{int} +H_{nn}
\label{Hubbard}
\ee
where in terms of creation operators $c_{\vec R,s,\tau}^\dagger$ for electrons with spin and orbital polarizations $s$ and $\tau$ on site $\vec R$.  The band structure is given by the nearest-neighbor tight-binding model
\be
H_0 =-t \sum_{\langle \vec R,\vec R^\prime \rangle,s,\tau} \left[c_{\vec R,s,\tau}^\dagger c_{\vec R^\prime,s,\tau}+ {\rm H.C.}\right] . 
\ee
The most general on-site interactions (consistent with the above stated symmetries) can be expressed as a sum of the ``important'' interactions
\be
H_{int} = {U}\sum_{\vec R} [\hat n_{\vec R} -2]^2 - K \sum_{\vec R} \left[{\bf L}_{\vec R}\cdot  {\bf L}_{\vec R} - \delta { L}^z_{\vec R}{ L}^z_{\vec R} \right]
\ee
where $\delta$ tunes between X--Y and Ising-like characters, and additional interactions
\ba
H_{int}^\prime = \sum_{\vec R} \left[ {U_3}+ {U_4}{(\hat n_{\vec R} -2)}\right]{(\hat n_{\vec R} -1)(\hat n_{\vec R} -2)(\hat n_{\vec R} -3)} .
\nonumber
\ea
For simplicity, we ignore $H_{int}^\prime$, \emph{i.e.} we set $U_3 = U_4 = 0$. $H_{nn}$ is an additional further-neighbor interactions (assumed small) that we will introduce below.  Here
\be
\hat n_{\vec R} =\sum_{\tau,s} c^\dagger_{\vec R,s,\tau} c_{\vec R,s,\tau},
\ee
and the orbital pseudo-spin
\be
\vec L_{\vec R} = \frac 1 2\sum_{\tau,\tau^\prime,s} c^\dagger_{\vec R,s,\tau} \vec \sigma_{\tau,\tau^\prime}c_{\vec R,s,\tau^\prime}.
\ee

To better appreciate the significance of $K$, consider the single site problem with two electrons:
For $K>0$, the spin triplet states  have higher energy than the singlet, in violation of  Hund's first rule.\footnote{We have not included an explicit $J\vec S \cdot \vec S$ coupling, as this can be absorbed as a shift of $U$ and $K$.}  Indeed, in the following we will assume $K>0$ which, as already indicated, involves the non-trivial effects of high energy degrees of freedom that have been integrated out.  %For simplicity, we have neglected other terms, \emph{e.g.} multi-orbital pair hopping.

The assumption that further neighbor interactions, $H_{nn}$, are relatively small deserves comment as well.  Of course, Coulomb interactions are long-ranged, but in the devices in question, there is a metallic gate separated from the bilayer by a distance comparable to the size of a unit cell in the moir\'e pattern.  We  invoke this as the justification for assuming that further neighbor interactions are weak, and can mostly be neglected.  
Since  the bare interactions between electrons are strongly repulsive, it is likely that $U> 0$, although possibly, due to the effects of electron-phonon coupling and other correlation effects, it may be smaller than naive estimates  suggest, and we will even consider the case in which it is less than $K$.

We also include in our model a (weak) nearest-neighbor interaction %between orbital spins
\be
H_{nn} =- \widetilde K \sum_{\langle \vec R,\vec R^\prime \rangle} {\bf L}_{\vec R}\cdot  {\bf L}_{\vec R^\prime}, 
\ee
and take $K \gg \widetilde K>0$. The latter condition implies that $H_{nn}$ favors orbital ferromagnetism, as would be expected for a direct exchange interaction.
Because the triangular lattice is non-bipartite, the sign of $t$ is significant.  We will take $t<0$ on a phenomenological basis to reproduce the observed sense of the particle-hole asymmetry about $\overline{n}_h=2$, and will at some places consider the effects of more complex band-structures, {\it i.e.} the effects of further neighbor hopping.

\section{Mean Field Theory}

For weak coupling, a reliable solution can be obtained using Hartree-Fock/BCS mean field theory.  In this limit, the only generic instability is to superconductivity.  If we nevertheless apply the same approach for intermediate couplings, the results -- while not well justified -- are highly suggestive.  Because we have dominantly on-site interactions, the only possible order parameters are on-site.  If we assume translation symmetry is unbroken, then these consist of orbital or spin ferromagnetism and superconductivity involving on-site pairing.  Since we are assuming $K>0$, spin ferromagnetism can be neglected and of the possible superconducting channels,  only orbital pseudo-spin-triplet spin-singlet pairing is favored.  To simplify the equations, in this section we will take $\widetilde K=0$.

To implement this approach, we define  a trial Hamiltonian $H_{tr}$
(where henceforth summation over spin and pseudo-spin indices, $s$ and $\tau$ are implicit):
\begin{equation}
\begin{split}
&H_{tr} =  \sum_{k}  \epsilon_k c^\dagger_{k, s, \tau} c_{k, s, \tau} - \vec{h} \cdot \sum_{\vec R} \vec{L}_{\vec R} \\ 
&+ \sum_{\vec R}  \left(  c^\dagger_{\vec R, \uparrow, \tau} \big[ ( \vec{d} \cdot \vec{\sigma} ) i \sigma_y \big]_{\tau, \tau'} c^\dagger_{\vec R, \downarrow, \tau'} + \text{H.C.} \right)
\end{split}
\end{equation}
where $\vec{h}$, $\vec{d}$, and $\epsilon_k$ are variational parameters corresponding to orbital ferromagnetism, triplet superconductivity, and band structure, respectively. Under time-reversal $\Theta$, $\vec h_\parallel \rightarrow \vec h_\parallel$, $h_z \rightarrow -h_z$, $\vec d_\parallel \rightarrow -\vec d_\parallel$, and $d_z \rightarrow d_z$, where $\vec h_\parallel$, $\vec d_\parallel$ are the X--Y components of $\vec{h}$ and $\vec{d}$, respectively.  
$F_{var} \equiv F_{tr} + \langle H - H_{tr} \rangle_{tr}$ is a variational upper bound to the free energy.  
The corresponding mean field equations are 
\begin{equation}    
h_j  = V^{F}_j\ \langle L_j(\vec R)\rangle
\label{eq:MF_h}
\end{equation}
\begin{equation}
d_j   =  - V^{sc}_j  \langle c_{\vec R \uparrow \tau} [ \sigma_j ( i \sigma_y ) ]^*_{\tau \tau'} c_{\vec R \downarrow \tau'} \rangle
\label{eq:MF_d}
\end{equation}
where $V^{F}_x=V^{F}_y =
  2U  + \frac{K}{2} ( 5 - \delta )  $, $V^{F}_z=  2U  + \frac{K}{2} ( 5 - 3\delta )  $, $V^{sc}_x=V^{sc}_y =
 U - \frac{K}{4} ( 1 - \delta  ) $, $V^{sc}_z= U - \frac{K}{4} ( 1 + \delta  ) $
and there is a shift in the chemical potential.
For weak coupling, the superconducting $T_c$ is determined by whichever component of $V^{sc}_j$ is most attractive: 
\begin{equation}
T_c \sim W e^{ -1/2N_0 |V^{sc}_j| }
\end{equation}
where $N_0$ is the density of states at the Fermi level %, $V = [ U - K ( 1 \pm \delta ) / 4 ] $ 
and $W$ is the bandwidth. If $\delta > 0$, then $\vec d =d\hat z$ and the superconducting state preserves time-reversal and $U(1) \times Z_2$ orbital symmetries.
If $\delta < 0$, then $\vec d$ lies in the X--Y plane, and may give rise to a rich variety of superconducting phases whose symmetries will be discussed at the end of this section.

Orbital ferromagnetism is possible at stronger coupling when the interactions exceed an appropriate Stoner criterion.  For $\delta >0$, the orbital moment $\langle \vec L \rangle$ lies in the X--Y plane and breaks the orbital $U(1)$ symmetry;  this is a form of nematic order.  For $\delta <0$, $\langle \vec L \rangle = L\hat z$;  time reversal symmetry is broken but the orbital $U(1)$ symmetry is preserved.

\begin{figure} 
  \includegraphics[width=3.4in]{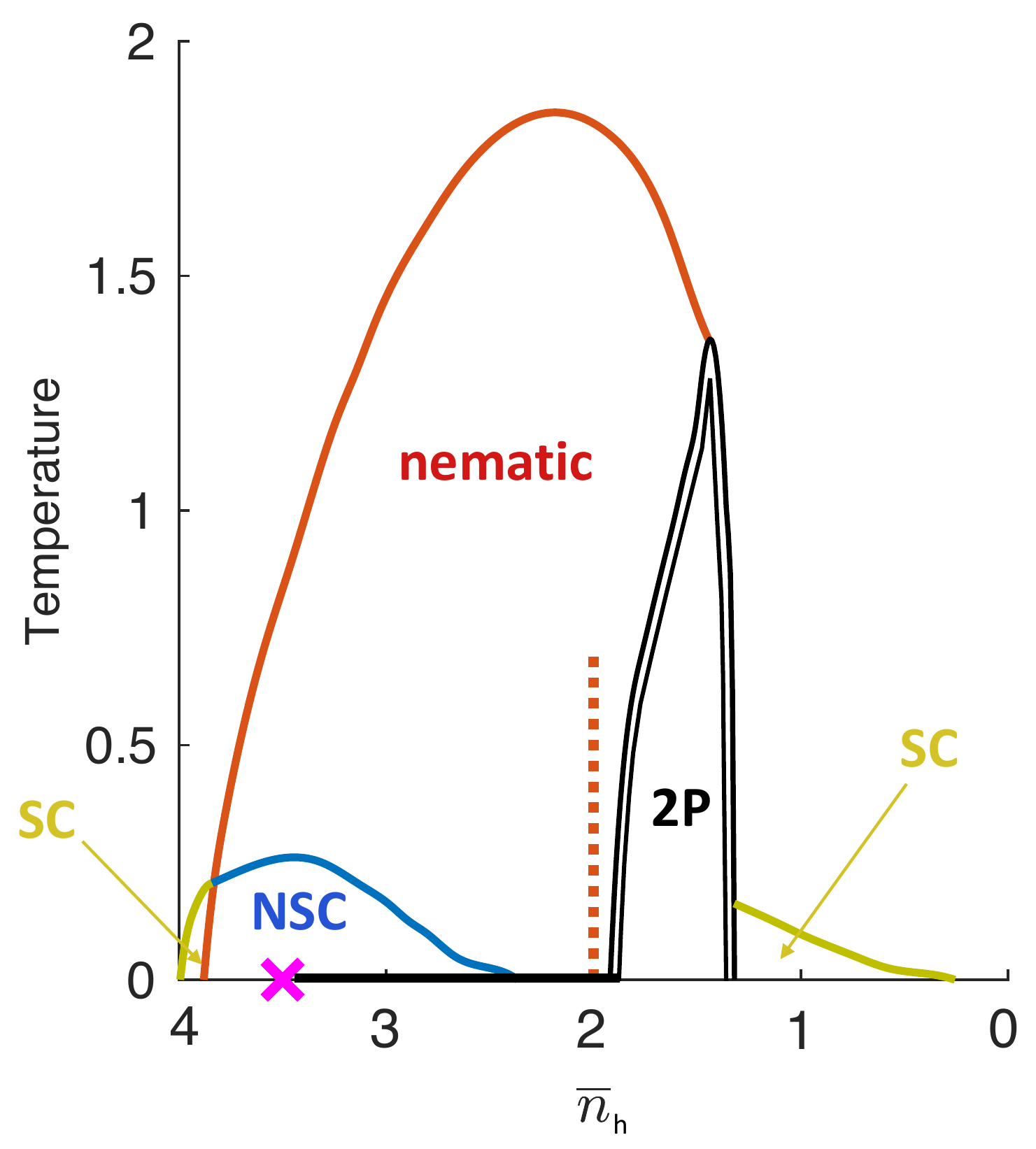}
  \caption{Mean-field phase diagram with parameters $U = 0$, $K = 4$, $t=-1$, $\delta = 0.3$ and $\widetilde K=0$ (temperature is measured in units of $|t|$). Note that the number of electrons per unit cell in the band immediately below charge neutrality, $\overline n_e$, is given by $4-\overline n_h$. Various broken symmetry phases are indicated in the diagram, where ``SC"  and ``NSC'' are, respectively,  a superconducting phase with only $d_z\neq 0$, and a nematic superconducting phase in which, in addition, the in-plane component of $\vec d$ perpendicular to $\vec h$ is non-zero.  
  The orbital ferromagnetic phase with $h_x \neq 0$ is fully polarized at $T=0$ for $n_{c1} \geq \overline n_h \geq n_{c2}$,
  as indicated by the solid black line.  This further implies that the ground state at $\overline{n}_h = 2$ is insulating. The dashed line at $\overline{n}_h = 2$ has a height corresponding to the insulating gap $|h_x|-W \approx 0.7$, where $W$ is the bandwidth. The regime labeled ``2P'' is a forbidden regime, where  two phases coexist macroscopically.  Solid colored lines represent continuous transitions and double lines discontinuous.  The cross at $\overline n_h=n_{c1}$ is a quantum critical point. }
  \label{fig:phasediagXY}
\end{figure}

\begin{figure} 
  \includegraphics[width=3.4in]{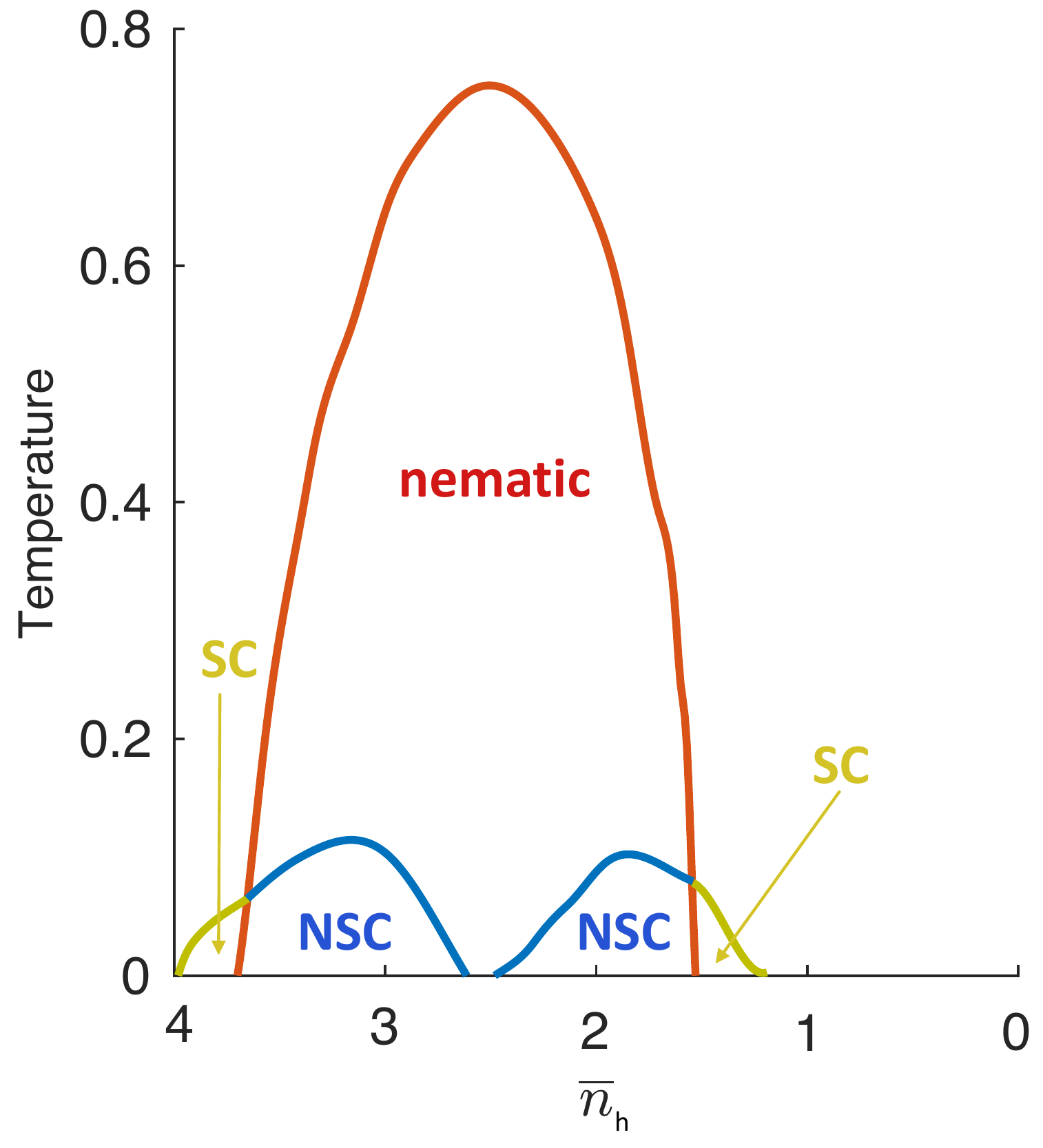}
  \caption{Mean-field phase diagram with parameters $U = 0$, $K = 2$, $t=-1$, $\delta = 0.3$, $\widetilde K=0$ and with a second neighbor hopping $t'=-0.4$. Symbols are as in Fig.~\ref{fig:phasediagXY}.  %The ground-state in this case is never fully polarized, 
  The state at $\overline n_h=2$ is conducting, and all transitions appear to be continuous.}
  \label{K2}
\end{figure}

The full mean-field solution of Eqs. \ref{eq:MF_h} and \ref{eq:MF_d} has been obtained numerically for the $\delta > 0$ case. Representative mean field phase diagrams are shown in Fig.~\ref{fig:phasediagXY}  for $U=0$, $K=4$, $t=-1$ and $\delta =0.3$,
and in Fig.~\ref{K2} for $U=0$, $K=2$, $t=-1$, $\delta=0.3$ and including a second neighbor hopping, $t^\prime = -0.4$.  The solid lines represent continuous phase boundaries while the double lines mark discontinuous transitions.  The broken symmetries are as indicated on the figures, where ``2P'' signifies a forbidden region of density in which  two-phase coexistence occurs, and ``SC'' and ``NSC'' refer, respectively, to superconducting and nematic superconducting phases.

The NSC exhibits a mixture of the pairing symmetries discussed above: the predominant channel has $d_z\neq 0$; however, the nematic order induces a non-zero X--Y component of  $\vec d$. Moreover, the X--Y components of $\vec d$ are related to the nematic order parameter $\vec h$: for $h_x \neq 0$, the NSC state has real $d_z$, imaginary $d_y$, and $d_x=0$.  The NSC therefore breaks gauge invariance and time-reversal simultaneously.
The pattern of order in the NSC phase can be understood by considering the symmetry allowed cubic terms that couple $\vec h$ and $\vec d$ in the Landau free energy $\mathcal{F}$,
\begin{equation} \label{eq:landau}
    \mathcal{F}(T) = \dots +  i a_1(T) \vec h_{\parallel} \cdot (\vec d^* \times \vec d)_{\parallel} + i a_2 (T) h_z (\vec d^* \times \vec d)_{z}, 
\end{equation}
%$+ a_3 (T)  \mathrm{Im}[d_z^*(\vec h_\parallel \cdot \vec d_\parallel)$ this term is redundant with the first term!
where $a_i(T)$ are temperature-dependent coefficients. Outside the nematic phase, the SC state has pure $d_z\neq 0$ order. 

For the $K=4$ case shown in Fig.~\ref{fig:phasediagXY}, 
the orbital pseudo-spin is fully polarized in the ground state for a range of $\overline n_h$ (indicated by the heavy black line in the figure) bounded from above by a mild quantum critical point at $\overline n_h = n_{c1}=3.5$ (marked by the cross) and from below by a first order point at $\overline n_h=n_{c2}\approx 1.9$. The system is insulating at $\overline n_h=2$.  If we were to suppress superconductivity (say by application of a magnetic field), then we would find  a (nematic) half-metallic orbital ferromagnet in this range of $\overline n_h$ (so long as $\overline n_h\neq 2$), and the transition that occurs at $\overline n_h=n_{c1}$ would be a Lifshitz transition.  
  
For the $K= 2$ case shown in Fig.~\ref{K2}, while there is a broad nematic phase, %but it
the order is not sufficiently strong to produce insulating behavior at $\overline{n}_h=2$.  Moreover, here we find that all the transitions appear to be continuous.
For substantially smaller $K$, there is no nematic phase at all.

If we were to include fluctuations beyond the mean-field treatment, since the nematic phase breaks a $U(1)$ symmetry, the ordered phase at finite temperature would presumably be replaced by a power law phase.  If crystal field effects were included that reduce the $U(1)$ to $C_6$, there could be a sequence of two transitions involving an ordered and a power law phase. Clearly all superconducting phases at non-zero $T$ would be replaced by phases with quasi-long-range-order. 

We now turn to address the possible phases for $\delta < 0$.  As mentioned above, we expect an orbital ferromagnetic phase with $h_z \neq 0$, which breaks time-reversal symmetry but preserves the orbital rotational symmetry. From Eq. \ref{eq:MF_d}, the leading superconducting instability leads to a non-zero $\vec d_\parallel$. The two-component nature of the order parameter can, by analogy to the case of Sr${}_2$RuO${}_4$,\cite{sigrist} lead to a number of different superconducting states. For simplicity, here we focus on states which can be reached from the non-superconducting phases by a continuous transition, {\it i.e.} those that break the minimal number of additional symmetries. Importantly, inside the orbital ferromagnetic dome, Eq. \ref{eq:landau} implies the relative phases of the $d_x$ and $d_y$ components lock in a way that preserves the orbital rotational symmetry. %Now, for the superconducting state outside the orbital ferromagnetic dome (with $\vec h = 0$), and assuming it can be continuously connected to the superconducting state under the orbital ferromagnetic dome, if we integrate out $h_z$ degree of freedom, we obtain a term in $\mathcal{F}$ that is proportional to $-|(\vec d^* \times \vec d)_{z}|^2$, and it diverges near the phase boundary. This again locks in the relative phase between $d_x$ and $d_y$, which implies that $U(1)$ symmetry is preserved. 
The different phases and corresponding broken symmetries that follow from Eq. \ref{eq:landau} are shown in Table 1 for both signs of $\delta$.    

\begin{table}
\begin{center}
    \begin{tabular}{ | c | c | c |}
    \hline
     & $\qquad  \text{X--Y-like} \qquad$ &  $\qquad \text{Ising-like} \qquad$ \\ \hline
    Normal ($\vec{h} \neq 0$)  & $\big[{C_6}\big ]$, $\Theta$, $C_2$ & $\big[{\Theta}\big]$, $C_6$  \\ \hline
    SC ($\vec h \neq 0$)  & $\big[{C_6}$, ${\Theta}\big]$, $C_2$ & $\big[{\Theta}\big]$, ${C_6}$   \\ \hline
    SC ($\vec h$ = 0) & %$\slashed{\Theta}$, $\slashed{C_6}$, $C_2$   
     $C_6$, $\Theta$ & $\big[ \Theta \big]$, $C_6$  \\ 
    \hline
    \end{tabular}
    \caption{Broken (in square brackets) and unbroken symmetries for the normal ($\vec{h} \neq 0$) state and various superconducting phases for $\delta < 0$ (Ising-like) and $\delta > 0$ (X--Y-like). The two choices of irreducible representations for the orbitals, $p \pm ip$ and $d \pm id$, are identical under rotation and time-reversal symmetries. However, mirror symmetries are affected differently in the two cases, and they also depend on which mirror plane one is considering.  }%In the Ising case with $\vec h=0$, there are various possible patterns of broken symmetry.\cite{***} }
\end{center}
\end{table}

%%%%%%%%%%%%%%%%%%%%%%%%%%%%%%%%%%%%%%%
\section{Strong Coupling Limit}

In much the same way as the $t-J$ model can be derived as the strong-coupling limit of the Hubbard model, we can derive an effective model that captures the low energy physics of our two-band Hubbard model in the strong coupling limit where $t$ is small compared to the on-site interactions.  Here, we map the problem to a lattice gas of hard-core particles. Let $\overline n_e$ be the number of electrons per unit cell in the band immediately below charge neutrality, $\overline n_e = 4-\overline n_h$. 
For compactness we consider the case in which $\delta$ is small. We further restrict our attention to $U> -K/2$ and the range $0\leq \overline n_e \leq 2$, but the case $2\leq \overline n_e \leq 4$ can be obtained from these results by a straightforward particle-hole transformation, which involves taking $t\rightarrow-t$. For $U > -K/2$, two doubly occupied sites have lower energy than either one singly and one triply occupied sites, or one quadruply occupied and one empty sites. Given this, for $0 \leq \overline n_e \leq 2$, we can project out the triply and quadruply occupied states.
Now 
\ba
H_{int} = \sum_{\vec R} \Big[\varepsilon_b b^\dagger_{\vec R,m}b_{\vec R, m} 
+\varepsilon_a a^\dagger_{\vec R,\tau,s}a_{\vec R,\tau,s}
%+{\bar a}^\dagger_{\vec R,\tau,s}{\bar a}_{\vec R,\tau,s}
 +4U \hat \nu_{\vec R} %+ q_{\vec R}^\dagger q_{\vec R}
\Big] 
\ea
where $\varepsilon_b\equiv -2K$, $\varepsilon_a\equiv U-3K/4$, and there is  a hard-core constraint that $\hat \nu_{\vec R}\geq 0$ for all $\vec R$, where $\hat \nu_{\vec R}$ is the density of empty sites,
\be
\hat \nu_{\vec R}=1- b^\dagger_{\vec R, m}b_{\vec R, m}-a^\dagger_{\vec R,\tau,s}a_{\vec R,\tau,s},
\ee
 and  the  electron density is $ 2-\hat \nu_{\vec R}$.
Here we represent doubly occupied sites as occupied by  a spinless $L=1$ boson with $L_z=m=-1$, 0, 1, with creation operator  $b^\dagger_{\vec R, m}$, while singly occupied sites are represented by $S=1/2$, $L=1/2$ fermions, with creation operator $a^\dagger_{\vec R,\tau,s}$. 

It is then easy to see that for $\overline n_e %\equiv \langle \hat n_{\vec R} \rangle 
= 2$ there is a Mott insulating state with one boson per site.  For  $U> K/4$, in the range $1 < \overline n_e <2$, the system is a mixture of $\langle a^\dagger_{\vec R,\tau,s}a_{\vec R,\tau,s}\rangle = (2-\overline n_e)$ fermions and $\langle b^\dagger_{\vec R, m}b_{\vec R, m}\rangle =\overline n_e -1$ bosons, while for $0 < \overline n_e< 1$, it contains a concentration $\langle a^\dagger_{\vec R,\tau,s}a_{\vec R,\tau,s}\rangle = \overline n_e$ fermions and no bosons.  On the other hand for $K/4 > U > -K/2$, in the entire range $0<\overline n_e < 2$ the system consists purely of bosons with 
$\langle b^\dagger_{\vec R, m}b_{\vec R, m}\rangle =\overline n_e/2$.  

We now consider the effect of the additional terms in the Hamiltonian.  In the obvious way, $H_{nn}$ can be rewritten as a nearest-neighbor orbital pseudo-spin dependent interaction between the bosons and fermions.  To first order in $t$ there are two terms generated:  a hopping term for fermions between  nearest-neighbor sites $\vec R$ and $\vec R^\prime$:
\be
-t\left[a^\dagger_{\vec R,\tau,s}a_{\vec R^\prime,\tau,s} + {\rm H.C.}\right], 
\ee
and a nearest-neighbor boson-fermion interchange:
\ba
&&t/2\left [(b^\dagger_{\vec R, m}\vec T_{m,m^\prime}b_{\vec R^\prime, m^\prime})\cdot( a^\dagger_{\vec R^\prime,\tau,s}\vec \sigma_{\tau,\tau^\prime} a_{\vec R,\tau^\prime,s}) +{\rm H.C.} \right]\nonumber \\
&&+t/2
\left [b^\dagger_{\vec R, m}b_{\vec R^\prime, m} a^\dagger_{\vec R^\prime,\tau,s} a_{\vec R,\tau,s} + {\rm H.C.}\right],
\label{exchange}
\ea
where $\vec T$ is the pseudo-spin-1 generators of rotations.

Finally, there are a variety of terms that are generated in higher order perturbation theory in powers of $t$, which are thus assumed to be small compared to terms of order $t$.  Of these, the two most  important are a renormalization of the interaction in $H_{nn}$ acting between two neighboring sites occupied by bosons,
\be
\widetilde K \to \widetilde K^{\text{eff}}= \widetilde K -\frac{4 t^2}{4U+5K}
\label{tildeK}
\ee
and a nearest-neighbor hopping term for the bosons with matrix element
\be
t^{\text{eff}} = \frac {t^2}{K-4U}.
\label{teff}
\ee
The first term is potentially important in the  state at $\overline n_e=2$;  here the symmetry of the Mott insulator changes from orbital ferromagnetism for $\widetilde K^{\text{eff}} >0$ to orbital antiferromagnetism %(or possibly some form of spin-liquid order) 
for $\widetilde K^{\text{eff}} < 0$.  The boson hopping term is small compared to the boson-fermion exchange, and so is important only when there are no fermions to induce delocalization of the bosons, {\it i.e.}  when $K/4 > U > -K/2$. Note that there is an interesting breakdown of perturbation theory near the upper end of this regime.

The physics of the resulting lattice gases is itself rich and interesting.  The case of orbital pseudo-spin-1 bosons (relevant for $K/4>U>-K/2$) is similar to the problem of spin-1 bosons that has been studied in the context of cold atomic gases.\cite{zhou_2003}  For ferromagnetic $\widetilde K>0$, this results in the existence of a spin polarized Bose condensate,\cite{ho_1998, machida_1998} while for antiferromagnetic $\widetilde K < 0$, there are multiple possible states including fractured condensates.\cite{ho_2000}  The more interesting case in which there is a boson-fermion mixture ($U > K/4$ and $1<\overline n_e < 2$) has an analogue in the problem of $^3$He-$^4$He mixtures.\cite{blumeemerygriffith}  Various forms of superfluid states as well as phase separated states are possible depending on parameters.

One feature of the strong-coupling limit ($U\to \infty$) that is particularly striking is a strong asymmetry in the orbital-ferromagnetism between $\overline n_e <2$ and $\overline n_e > 2$.  This can be seen by studying the analogue of the Nagaoka problem\cite{nagaoka} -- either one doped electron or one doped hole relative to $\overline n_e=2$.  For $t<0$, Nagaoka's theorem applies to one hole, but not to one electron.  More detailed analysis\cite{dmrg,us} shows that dilute concentrations of doped holes make an effective ferromagnetic contribution to $\widetilde K^{\text{eff}}$ proportional to $t(2-\overline n_e)$ while doped electrons make a corresponding antiferromagnetic contribution.  While this result is restricted to infinite $U$ and vanishing doping density, in the next section we will show that the basic physics is much more robust. 
\footnote{If $U$ is sufficiently large, it is possible to find insulating states at $n_h=1$ and $n_h=3$, as is clear from the strong coupling analysis.  In this limit, each site has a spin 1/2 and an orbital spin 1/2, the ordering of which is described by some sort of effective model.  If we continue to stick to states that do not break translation symmetry, then the states in question will exhibit both orbital and spin ferromagnetism.  Consequently, nearby metallic phases would be nematic if the orbital ferromagnetism has X--Y character, and would exhibit an anomalous  Hall effect in the Ising case. This is, however, a very different parameter range -- reflecting very different microscopic physics -- than we have considered in the body of the paper.}

\section{Exact Diagonalization}

\begin{figure}
    \centering
    \includegraphics[width=0.48\textwidth]{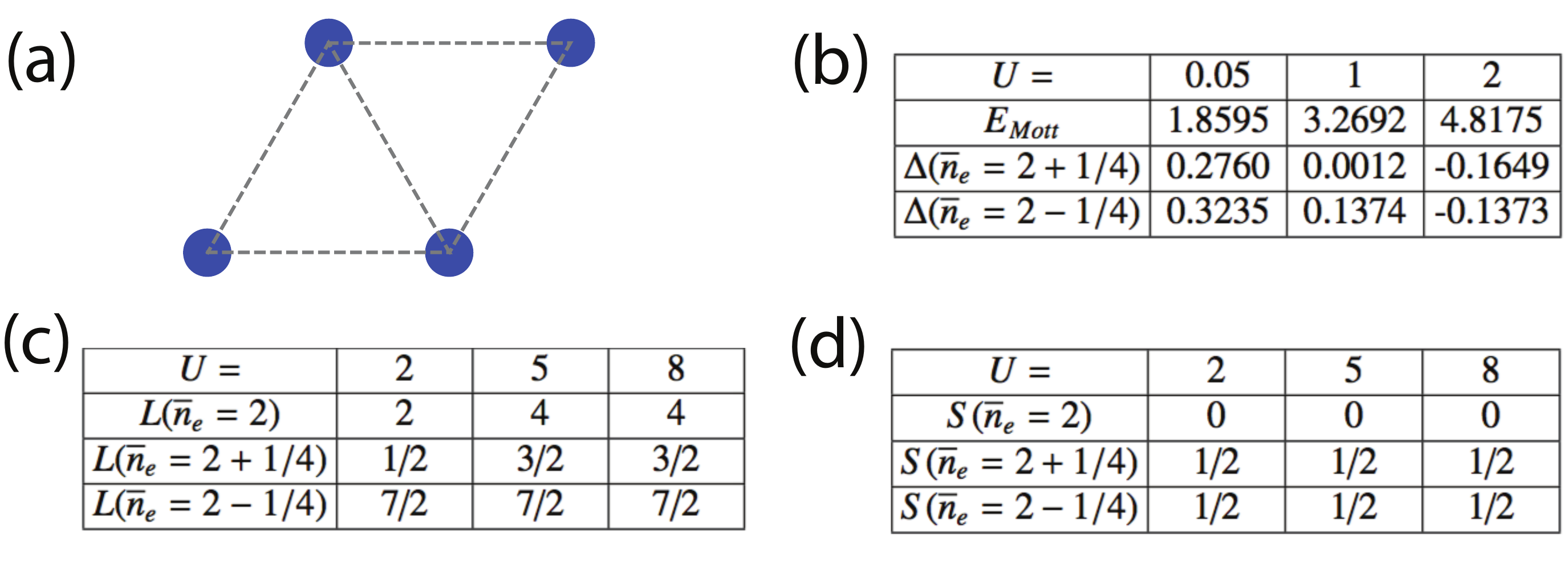}
    \caption{(a) The four site system that we treat by exact diagonalization using the Hamiltonian in Eq.~\ref{Hubbard}. We take $K=2$, $t=-1$, $\delta =0$, $\widetilde K = 0.2$, and various values of $U$. (b) The quantities $E_{Mott}$ and $\Delta(\overline{n}_e)$ are defined in Eqs. \ref{mott} and \ref{sc}. (c) The total orbital pseudo-spin of the ground state for various $\overline{n}_e$. (d) The total spin of the ground state for various $\overline{n}_e$.}
    \label{fig:ED}
\end{figure}

We have studied the model in Eq.~\ref{Hubbard} on the  $N=4$ site cluster shown in Fig. \ref{fig:ED} for various values of $U$.  We have taken units such that $t=-1$ and have arbitrarily taken $K=2$, $\widetilde K=0.2 $, and $\delta=0$.  (This latter condition implies an enlarged $SU(2)$ orbital symmetry and therefore $L$ is conserved.)  This is a  small system, but large enough to allow us to make crude estimates of interesting properties of the system at intermediate coupling, where analytic approaches fail.  The results with various numbers of electrons are representative of the properties of a larger system with values of $\overline n_e = 2, \ 2\pm 1/4,\ 2\pm 1/2$, etc.  We  compute the total orbital pseudo-spin $L$;  if this takes its maximal allowed value, $L(\overline n_e)=L_{max}(\overline n_e)=N\overline n_e/2$, then this is suggestive that the larger system will exhibit a fully polarized orbital ordered state, while there is likely no orbital ferromagnetism in the thermodynamic limit if $L$  takes its minimal value $L(\overline n_e)=L_{min}(\overline n_e)$ where $L_{min}(\overline n_e)=0$ or $1/2$ depending on whether $N\overline n_e$ is even or odd.  From the ground-state energy as a function of electron number, $E(N_{el})$,  we  define two particularly interesting  ``gaps''
\be
E_{Mott}\equiv E(2N+1) + E(2N-1) - 2 E(2N)
\label{mott}
\ee
and, for $\overline n_e \neq 1,\ 2,$ or 3 and under conditions $N\overline n_e=$ odd,
\be
\Delta(\overline n_e) \equiv [2E({\overline n_e N})-E({\overline n_e N +1}) - E({\overline n_e N-1})]/2
\label{sc}
\ee
These quantities are defined such that if there were an insulating state at $\overline n_e=2$, then  $E_{Mott}$ would indeed approach the Mott gap in the limit $N\to \infty$ and if the doped system were to have a nodeless superconducting gap, then $\Delta(\overline n_e)$ would approach the value of the minimal gap in the same limit.  
Some results for $L(\overline n_e)$, $S(\overline n_e)$, $E_{Mott}$, and $\Delta(\overline n_e)$  are reported in Fig.~\ref{fig:ED}. In the case $\overline n_e < 2$, the system is a fully polarized  orbital ferromagnet, while in the case $\overline n_e > 2$, the  orbital pseudo-spin is far from being fully polarized. There is a superconducting gap $\Delta(\overline{n}_e)>0$ when $U$ is small that is destroyed for larger $U$.

\section{Exact results for $U=K/4$}

Under the special condition $U=K/4$ with $\delta =0$ and $\widetilde K=0$, the model defined in Eq.~\ref{Hubbard} becomes effectively non-interacting in the sector of Hilbert space with maximal orbital pseudo-spin \footnote{Some of these results can be extended to  $\delta<0$.}.
The model itself is still strongly interacting; it is just in this sector that the interactions effectively vanish.  It still depends on $\overline n_e$ and $U$ whether or not the ground-state is a fully polarized orbital ferromagnet. When this is the case, then if $\overline n_e=2$ the system is insulating while for $\overline n_e \neq 2$ it forms a half-metallic orbital ferromagnetic Fermi gas.  

It is also interesting to consider what happens upon perturbing about this solvable line.  For small $\widetilde K >0$ or $U < K/4$, there are effective weak attractive interactions induced between like-orbital pseudo-spin electrons, and this in turn leads -- via the usual BCS mean field analysis -- to a spin singlet orbital pseudo-spin triplet superconducting state of the sort discussed in the mean-field analysis above.  Otherwise, the effective interactions are repulsive and any superconducting state that arises will have unconventional pairing and will arise by a version of the Kohn-Luttinger mechanism or through the exchange of a collective boson leading to a parametrically lower $T_c$.

\section{Relation to other work}

Not surprisingly, the discovery of superconductivity in twisted bilayer graphene has produced a flurry of ``rapid response'' theories,\cite{fu_2018, ramires_2018, volovik_2018, roy_2018, scalettar_2018, baskaran_2018, phillips_2018} of which the present paper is one. It is important to stress that  the experimental situation is still evolving, so  the applicability of any theoretical proposal is presently difficult to judge.  

Our perspective differs from that of the other papers of which we are aware in several important ways.  Many of these works make an implicit analogy with the cuprates in identifying the insulating behavior at $\overline{n}_h=2$ as ``Mott insulating,'' and looking for possible mechanisms of unconventional pairing in which the pair-wave function vanishes for two electrons on the same site.  These papers all envisage the dominant interaction to be a strongly repulsive Hubbard $U$, and where weaker interactions are considered, they are such as to favor Hund's (first) rule states with maximal spin.  In contrast, we have explored the possibility that a combination of correlation effects (involving bands that have been integrated out) and electron-phonon effects lead to a reduction of $U$ and violations of Hund's rule, analogous to the  situation that is believed to apply in alkali doped C$_{60}$.  We have proposed a large role for orbital pseudo-spin ferromagnetic ordering, including identifying this broken symmetry as the cause of the insulating phase.  Moreover, we have suggested that the large asymmetry in the behavior of the quantum oscillations for $4>\overline{n}_h>2$ versus $2 > \overline{n}_h > 0$ is associated with Nagaoka-type stabilization of the orbital ferromagnetism for one sign of doping but not the other.
We have also found a superconducting state that  is largely conventional with on-site spin-singlet pairing (albeit with possibly interesting orbital pseudo-spin structure). 

There are also some important differences and similarities in the models considered.  The model we study is in large part the same as that introduced in Ref. \onlinecite{xu_2018}, albeit they considered it in a very different range of parameters (where, for example, Hund's first rule is obeyed).  As mentioned earlier, several papers\cite{senthil_2018,fu_2018} have noted that there are significant theoretical barriers to a direct route from the nodal band structure of the individual graphene sheets to the effective two-orbital model we have studied.  We have no disagreement with this conclusion -- we view our model as a
phenomenological construct.

 An ambitious approach to connecting the microscopics to the observed phenomena is reported in Ref. \onlinecite{senthil_2018}.  In contrast with our paper, this discussion purports to deal  with the strong repulsive interactions between the conduction electrons without invoking the high energy renormalizations discussed above.  On the other hand, from a correspondingly more complex analysis, they also identify the Mott insulating phase as 
 being the same sort of nematic as we have found, and have suggested that the superconducting state has a spin-singlet s-wave gap structure.  The principal mechanism of pairing in their work is exchange of collective fluctuations;  we, too, think this could play a role,\cite{pnas} but since we already obtain pairing at the mean-field level, we imagine that the main role of such fluctuations is to enhance  $T_c$ in the neighborhood of the nematic quantum critical point.\cite{lederer,doug}

\section{Relation to Experiment in Twisted Bilayer Graphene}

We have introduced a  two-band Hubbard-like model on a triangular lattice as the simplest lattice scale model with enough degrees of freedom to account for certain salient features of present experimental observations\cite{jarillo-INS_2018, jarillo-SC_2018} in twisted bilayer graphene, doped with a concentration $\overline{n}$ of holes relative to charge-neutrality.  Even this stripped down model is complex and has a large variety of possible ordered phases and phase diagram topologies as a function of the various interactions, some of which we have elucidated.  In this final section, we summarize some of these results and their possible relevance to bilayer graphene. 

In the strong coupling limit, we find a Mott insulating state at $\overline n_h=2$ in the sense that the insulating behavior onsets at a high temperature and exhibits a  gap associated directly with the on-site interactions.  The scale of this gap is large compared to inter-site couplings, and hence to the temperatures at which any ordering phenomena occur.  This behavior is analogous to what is seen in the cuprates, where the gap in the insulating state of the undoped parent compounds is of order 2 eV, while all ordering phenomena occur at around room temperature or below.  Such behavior is {\em not} seen in twisted bilayer graphene where the resistivity changes from a high temperature metallic behavior to a low temperature insulating behavior at 4 K, which is just over a factor of 2 larger than the largest superconducting $T_c$.  We thus conclude that it is likely that the insulating behavior in bilayer graphene is associated with a broken symmetry state, and should be associated with the intermediate coupling regime of parameters in our model.

One attractive candidate for the insulating state is a fully polarized orbital ferromagnet.  Depending on the sign of $\delta$, this state can either have X--Y character -- in which case it is some sort of electron nematic state -- or Ising character -- in which case it breaks time-reversal symmetry and would %likely lead to 
result in a zero-field anomalous Hall effect and an offset in the quantum oscillations in metallic state away from $n_h = 2$. 
%That no such signature is seen in experiment favors $\delta>0$ and a nematic state.  
While a fully polarized nematic state occurs naturally in the strong-coupling limit, it is unclear down to what level of couplings it survives.  Nonetheless, we do find such an insulating state in our mean-field treatment at intermediate values of $K \sim E_F$.  (See Fig.~\ref{fig:phasediagXY}.)  %The insulating behavior requires orbital and spin ferromagnetism or broken translational symmetry

One interesting feature of the nematic state that arises naturally in the strong-coupling limit  is an  intrinsic particle-hole asymmetry  with respect to doping away from $\overline n_h=2$.  Doping to $\overline{n}_h > 2$ produces an additional tendency to orbital ferromagnetism (induces an effective increase in the magnitude of $\widetilde K$) while $\overline{n}_h < 2$ favors orbital antiferromagnetism (tends to make $\widetilde K$ negative).  Needless to say, orbital ferromagnetism is entirely unfrustrated even on a triangular lattice, while orbital antiferromagnetism is highly frustrated and likely has a much reduced ordering temperature -- if it orders at all.  The same particle-hole asymmetry  around $\overline{n}_h=2$ appears as a robust feature in the exact diagonalization analysis for intermediate couplings, and can occur at mean-field level as well, provided the density of states is sufficiently  asymmetric, as can be seen in Figs.~\ref{fig:phasediagXY} and \ref{K2}.

For $\overline n_h > 2$, if we assume that the system is a fully  polarized orbital ferromagnetic Fermi liquid, then from Luttinger's theorem it follows that the area enclosed by the Fermi surface is $A =A_0\ (\overline n_h-2)/2 $ mod $A_0$, where $A_0$ is the area of the Brillouin zone.  By contrast, if for a range of $\overline{n}_h<2$ the system forms an unpolarized Fermi liquid, then the area enclosed by the Fermi surface is $A= A_0\ n_h/4 $.  These expressions are loosely consistent with what has been inferred from quantum oscillation experiments in twisted bilayer graphene. 

The conjectured nematicity -- if it exists in the ground state -- would necessarily onset at a finite transition temperature, $T_{F}(\overline n_h)$.  In a homogeneous system, this would imply the existence of singular temperature dependences of various measured quantities.  No such singular behavior has been observed.  However, there are reasons to believe that the electronic structure of currently available materials is somewhat inhomogeneous, which would lead to a rounding of such singularities.  We would thus tentatively like to associate the temperature of the observed metal-insulator crossover with this phase transition.  This temperature appears to be maximal around $\overline n_h=2$ and to drop smoothly with increasing $\overline{n}_h$ extrapolating to 0 at a critical doping $\overline n_{h,c} \approx 2.4$.  In this interpretation, $n_c$ is associated with a nematic quantum critical point.  The fact that this critical point roughly coincides with the hole doping at which the maximal $T_c$ occurs, invites analogy with the phase diagrams of the Fe-based high temperature superconductors, where nematic quantum critical fluctuations have been conjectured to enhance $T_c$.\cite{lederer,doug,fisher}  

Turning now to the primary mechanism of superconductivity:  The same interactions that promote the nematic order  also give rise to superconductivity.  We find   spin-0 pseudo-spin-1 pairing with a substantial on-site component. In this sense, the superconducting state reflects the existence of  attractive interactions.   We  suggest that a good analogy exists with another C-based superconductor, alkali doped C$_{60}$, where effective intra-molecular attractive interactions are generated by a (still somewhat unresolved) combination of intra-molecular strong correlation effects\cite{sudipandme,jiangandme} and electron-phonon couplings.\cite{assa, dynamicJT,c601,c602,c603,c604,granath_2002,granath_2003}  
In the calculations we have performed, this physics is encoded in the assumed values of $U$ (assumed to be relatively small) and $K$ (assumed to be positive).  
This is in contrast with the situation in unconventional superconductors where the interactions are strongly repulsive, and significant (albeit often very short-range-correlated) antiferromagnetic correlations often coexist with superconductivity.

The spin-singlet character of the superconducting order is consistent with the relatively small value of the in-plane critical field observed in experiment.  At least the form of the pair wave-function should be clarified once some spectroscopic probes of the gap structure -- possibly tunnelling experiments -- are carried out.  Somewhat in analogy with the situation in $^3$He, there is a complex order parameter space associated with the pseudo-spin-1 character of the pairing which can lead to interesting textures and topology.  A tendency to phase separation of the sort shown in Fig.~\ref{fig:phasediagXY} could lead to an enhanced susceptibility to the formation of electronically inhomogeneous states, even in otherwise relatively homogenous samples.

\begin{acknowledgments}  We acknowledge useful discussions with  E. Berg, I. Esterlis, Y. Gannot, H. Guo, P. Jarillo-Herrero, L. Levitov, C. H. Mousatov, R. Scalettar, T. Senthil, A. Vishwanath, and C. Wu.
This work was supported in part by the Department of Energy, Office of Basic Energy Sciences, under contract no. DE-AC02- 76SF00515 at Stanford (JFD, SAK, YS, XQS and CW). YS was also supported by the Zuckerman STEM Leadership Program.
\end{acknowledgments}

%%%%%%%%%%%%%%%%%%%%%%%%%%%%%%%%%%%%%%%
\begin{appendix}

\end{appendix}


\begin{thebibliography}{99}
\bibitem{us}  We have shown by an explicit variational calculation that in the $U \to \infty$ limit of the present model, the ground-state of a finite size system with one doped hole is a fully polarized orbital ferromagnet, consistent with Nagaoka's theorem, while for one doped electron an antiferromagnetic state has lower energy.
\end{thebibliography}
\end{document}